\documentclass{ws-procs9x6}

\newcommand{\refeq}[1]{(\ref{#1})}
\def\etal {{\it et al.}}

\def\al{\alpha}

\def\de{\delta}
\def\ep{\epsilon}

\def\ka{\kappa}
\def\la{\lambda}

\def\vs{\varsigma}

\def\om{\omega}

\def\mn{{\mu\nu}}

\def\prt{\partial}
\def\cl{{\cal L}}
\def\vev#1{\langle {#1}\rangle}

\def\fr#1#2{{{#1} \over {#2}}}
\def\frac#1#2{{{#1} \over {#2}}}

\def\half{{\textstyle{1\over 2}}}
\def\lsim{\mathrel{\rlap{\lower4pt\hbox{\hskip1pt$\sim$}}
    \raise1pt\hbox{$<$}}}
\def\gsim{\mathrel{\rlap{\lower4pt\hbox{\hskip1pt$\sim$}}
    \raise1pt\hbox{$>$}}}
\def\sqr#1#2{{\vcenter{\vbox{\hrule height.#2pt
         \hbox{\vrule width.#2pt height#1pt \kern#1pt
         \vrule width.#2pt}
         \hrule height.#2pt}}}}
\newcommand{\beq}{\begin{equation}}
\newcommand{\eeq}{\end{equation}}
\newcommand{\bea}{\begin{eqnarray}}
\newcommand{\eea}{\end{eqnarray}}

\def\mbf#1{\mbox{\boldmath$#1$}}
\def\syjm#1#2{\phantom{}_{#1}Y_{#2}}

\def\etal{{\it et al.}}

\def\kf{\hat k_F}
\def\kaf{\hat k_{AF}}

\def\kfd#1{k_{F}^{(#1)}}

\def\kafd#1{k_{AF}^{(#1)}}
\def\kde{\hat\ka_{DE}}
\def\kdb{\hat\ka_{DB}}
\def\khe{\hat\ka_{HE}}
\def\khb{\hat\ka_{HB}}

\def\voc{\mathrel{\rlap{\lower0pt\hbox{\hskip1pt{$c$}}}
    \raise3pt\hbox{$\neg$}}}
\def\vok{\mathrel{\rlap{\lower0pt\hbox{\hskip1pt{$k$}}}
    \raise6pt\hbox{$\neg$}}}

\def\sk#1#2#3{#1^{(#2)}_{#3}}

\def\kafoE{\sk{(\vok_{AF}^{(d)})}{1E}{njm}}

\def\kftE{\sk{(\vok_F^{(d)})}{2E}{njm}}
\def\kfoE{\sk{(\vok_F^{(d)})}{1E}{njm}}

\def\kftB{\sk{(\vok_F^{(d)})}{2B}{njm}}

\def\kjm#1#2#3{k^{(#1)}_{(#2)#3}}
\def\cjm#1#2#3{c^{(#1)}_{(#2)#3}}
\def\kI{\cjm{d}{I}{jm}}

\def\kE{\kjm{d}{E}{jm}}
\def\kB{\kjm{d}{B}{jm}}
\def\kV{\kjm{d}{V}{jm}}

\def\cftzE{\sk{(\voc_F^{(d)})}{0E}{njm}}
\def\kftzE{\sk{(\vok_F^{(d)})}{0E}{njm}}
\def\kftoB{\sk{(\vok_F^{(d)})}{1B}{njm}}
\def\kaftzB{\sk{(\vok_{AF}^{(d)})}{0B}{njm}}
\def\kaftoB{\sk{(\vok_{AF}^{(d)})}{1B}{njm}}

\begin{document}

\title{HIGHER-ORDER LORENTZ VIOLATIONS IN ELECTRODYNAMICS}

\author{Matthew Mewes}

\address{
  Department of Physics and Astronomy, Swarthmore College\\
  Swarthmore, PA 19081, U.S.A.
}

\begin{abstract}
The Standard-Model Extension (SME) provides
a theoretical framework for tests of
Lorentz invariance.
To date, most studies have focused
on the minimal SME,
which restricts attention to operators of
renormalizable dimension.
Here, we review recent studies involving the
nonrenormalizable photon sector of the SME.
\end{abstract}

\bodymatter

\section{Introduction}

Experiments involving photons have
led to some of the most precise tests
of Lorentz invariance.\cite{datatables}
These include tests involving
resonant cavities,\cite{cavities}
cosmic birefringence,\cite{km_prd09,km_apjl,bire}
and  accelerators.\cite{accel}
Most searches for Lorentz violation in
photons have been analyzed using the
minimal version of the
Standard-Model Extension (SME).\cite{ck}
The SME provides the theoretical
foundation for Lorentz tests
involving any of the particles
in the Standard Model of particle physics
and gravity.
In its simplest form,
the minimal Standard-Model Extension (mSME)
restricts attention to operators that
obey the usual spacetime-translational
and gauge symmetries and
restricts attention to operators of
renormalizable dimension, $d=3,4$.

The photon sector of the mSME is given by
\beq
\cl=-\tfrac14 F_{\mu\nu}F^{\mu\nu}
+\half (k_{AF})^\ka\ep_{\ka\la\mu\nu}A^\la F^{\mu\nu}
-\tfrac14 (k_F)_{\ka\la\mu\nu} F^{\ka\la}F^{\mu\nu} \ .
\label{mSME}
\eeq
In addition to the usual Maxwell term,
this lagrangian includes two Lorentz-violating terms,
one for CPT-violating dimension-3
operators with coefficients
$(k_{AF})^\ka$
and another for 
CPT-conserving dimension-4 operators
with coefficients
$(k_F)_{\ka\la\mu\nu}$.
The renormalizable condition greatly
restricts the number of
Lorentz-violating operators.
There are a total of four independent
CPT-odd coefficients and nineteen
independent CPT-even coefficients.
The photon sector of the full SME includes
many more terms.
Here, we review a recent study of
nonrenormalizable operators in the photon
sector of the SME.\cite{km_prd09}

\section{Nonrenormalizable coefficients}

Allowing for operators of
arbitrary dimension,
but restricting attention to
those that maintain the usual
gauge invariance,
we arrive at a theory that 
is given by the Lagrange density
\beq
\cl = -\tfrac14 F_\mn F^\mn 
+ \tfrac12\ep^{\ka\la\mu\nu} A_\la (\kaf)_\ka F^{\mu\nu}
-\tfrac14 F_{\ka\la} (\kf)^{\ka\la\mu\nu} F_{\mu\nu} \ .
\label{SME}
\eeq
Here we only consider terms quadratic
in the photon field $A_\mu$,
leading to a linear theory.
The above equation is written
in a form that resembles the 
mSME expression \refeq{mSME}.
The key difference 
is that here the $\kaf$ and $\kf$
are operators that depend on
the 4-momentum $p_\mu = i\prt_\mu$.
The effects of these operators
mimic the effects of a permeable medium
whose activity depends on the
photon energy and momentum.
This introduces new frequency and
direction dependences that do not
arise in the mSME case.

The $\kaf$ and $\kf$ operators
can be expanded in $\prt_\mu$,
giving the expressions
\begin{align}
  (\kaf)_\ka&=\sum {(\kafd{d})_\ka}^{\al_1\ldots\al_{(d-3)}}
  \prt_{\al_1}\ldots\prt_{\al_{(d-3)}} \ , \notag \\
  (\kf)^{\ka\la\mu\nu}&=\sum (\kfd{d})^{\ka\la\mu\nu\al_1\ldots\al_{(d-4)}}
  \prt_{\al_1}\ldots\prt_{\al_{(d-4)}} \ ,
  \label{ks}
\end{align}
where we sum over the dimension $d$
of the associated operator.
The CPT-odd coefficients 
${(\kafd{d})_\ka}^{\al_1\ldots\al_{(d-3)}}$
are nonzero for odd $d\geq 3$.
The CPT-even coefficients 
$(\kfd{d})^{\ka\la\mu\nu\al_1\ldots\al_{(d-4)}}$
are nonzero for even $d\geq 4$.
The mSME case corresponds to $d=3,4$.
The number of coefficients for
a given dimension scales like $d^3$
for large $d$.

The large number of 
$\kafd{d}$ and $\kfd{d}$
coefficients and their relatively
complicated symmetries implies that
a systematic decomposition into a minimal
set of independent components is useful.
One decomposition uses
tensor spherical harmonics.
The idea is that each term in the sums
in Eq.\ \refeq{ks} takes
the form of a polynomial in
frequency $\om = p^0$
and momentum $\vec p$.
We can then expand these terms in
spin-weighted spherical harmonics
$\syjm{s}{jm}(\hat p)$.
The symmetries of the
$\kafd{d}$ and $\kfd{d}$ tensors
then impose constraints on the
spherical coefficients in the expansion.
These constraints can be used to find a
set of independent coefficients.

As an illustration,
the CPT-odd coefficients
split into four sets of spherical coefficients:
\beq
\kafd{d} \rightarrow \kV \oplus
\kaftzB \oplus \kaftoB \oplus \kafoE \ .
\eeq
The notation is chosen to provide
information about the properties
of the associated operators.
The index $d$ is the dimension of the operator.
The $j$ and $m$ are the usual spherical-harmonic indices
and determine behavior under rotations.
The index $n$ gives information concerning the
frequency/wavelength dependence of the operator.
The remaining indices, in parentheses,
give the helicity and parity of the operator.
Parity is labeled as either $E$-type, $(-1)^j$,
or $B$-type, $(-1)^{j+1}$.
Birefringent coefficients are labeled by $k$,
while $c$ denotes nonbirefringent coefficients.
Finally, the negation symbol $\neg$
is used to indicate vacuum-orthogonal 
coefficients - those that do not affect
light propagating in a vacuum,
at leading order.
Those that affect light appear without the $\neg$ symbol
and are referred to as vacuum coefficients.
The sets of spherical coefficients for Lorentz
violation for both CPT-odd and CPT-even operators
are summarized in Table \ref{table}.

\begin{table}[t]
  \tbl{Summary of spherical coefficients for Lorentz violation
    and their index ranges.}{
    \renewcommand{\arraystretch}{1.25}
    \begin{tabular}{c|c|c|c|c}
      coeff.\ & $d$ & $n$ & $j$ & number \\
      \hline
      \hline
      $\kI$ & even, $\geq 4$ &--& $0,1,\ldots, d-2$ & $(d-1)^2$   \\
      $\kE$ & even, $\geq 4$ &--& $2,3,\ldots, d-2$ & $(d-1)^2-4$ \\
      $\kB$ & even, $\geq 4$ &--& $2,3,\ldots, d-2$ & $(d-1)^2-4$ \\
      $\kV$ & odd,  $\geq 3$ &--& $0,1,\ldots, d-2$ & $(d-1)^2$   \\[4pt]
      \hline
      &&&&\\[-10pt]
      $\cftzE$  & even, $\geq 4$ 
      & $0,\ldots, d-4$ & $n,n-2,\ldots, \geq 0$    
      & $\frac{(d-1)(d-2)(d-3)}{6}$ \\
      $\kftzE$  & even, $\geq 6$ & $1,\ldots, d-4$   
      & $n,n-2,\ldots, \geq 0$
      & $\frac{(d-1)(d-2)(d-3)-6}{6}$ \\    
      $\kfoE$  & even, $\geq 6$ & $1,\ldots, d-4$ 
      & $n+1,n-1,\ldots, \geq 1$    
      & $\frac{(d-4)(d^2+d+3)}{6}$ \\
      $\kftE$  & even, $\geq 6$ & $2,\ldots, d-4$ 
      & $n,n-2,\ldots, \geq 2$    
      & $\frac{(d-4)(d^2-2d-9)}{6}$ \\
      $\kftoB$  & even, $\geq 6$ & $1,\ldots, d-4$   
      & $n,n-2,\ldots, \geq 1$    
      & $\frac{d(d-2)(d-4)}{6}$ \\    
      $\kftB$  & even, $\geq 6$ & $1,\ldots, d-4$ 
      & $n+1,n-1,\ldots, \geq 2$    
      & $\frac{(d+3)(d-2)(d-4)}{6}$ \\
      $\kaftzB$  & odd, $\geq 5$ & $0,\ldots, d-4$   
      & $n,n-2,\ldots, \geq 0$    
      & $\frac{(d-1)(d-2)(d-3)}{6}$ \\    
      $\kaftoB$  & odd, $\geq 5$ & $0,\ldots, d-4$   
      & $n+1,n-1,\ldots, \geq 1$    
      & $\frac{(d+1)(d-1)(d-3)}{6}$ \\    
      $\kafoE$ & odd,  $\geq 5$ & $1,\ldots, d-3$ 
      & $n,n-2,\ldots, \geq 1$    
      & $\frac{(d+1)(d-1)(d-3)}{6}$ \\
  \end{tabular}}
  \label{table}
\end{table}

\section{Astrophysical tests}

The split between vacuum and vacuum-orthogonal
coefficients is convenient because 
defects in the behavior of light
can be constrained with extreme precision
using sources at cosmological distances
$\sim 10^{42}$ GeV$^{-1}$.
The key equation for these tests is the
velocity of light in empty space,
\beq
v \simeq 1-\vs^0 \pm \sqrt{(\vs^1)^2+(\vs^2)^2+(\vs^3)^2} \ ,
\label{velocity}
\eeq
where
\begin{align}
  \vs^0 &= \sum E^{d-4} (-1)^j\, \syjm{0}{jm}(\mbf{\hat p})\, \kI \ ,\notag \\
  \vs^1\pm i \vs^2 &= \sum E^{d-4} (-1)^j\, \syjm{\pm2}{jm}(\mbf{\hat p})\big(\kE\mp i\kB\big) \ ,\notag \\
  \vs^3 &= \sum E^{d-4} (-1)^j\, \syjm{0}{jm}(\mbf{\hat p})\, \kV \ .
\end{align}
Several unconventional properties
result from these expressions.
First, the two signs in Eq.\ \refeq{velocity}
are associated with two different polarizations.
This leads to birefringence and an evolution
of polarization as light propagates.
Extremely tight constraints have been
placed on birefringence using light from
distant sources,
such as GRBs and the CMB.\cite{km_apjl,bire}
Lorentz violation also leads to direction
dependence in the speed of light.
All coefficients with $j\neq 0$ result
in anisotropies.
Additionally, we get an energy-dependent
speed when $d\neq 4$, giving dispersion.

Many of these effects are also present in
the mSME, but the new operators introduce
more complexity, such as direction dependence
with higher-order multipoles.
Dispersion results in the mSME case from $d=3$ coefficients,
but it is accompanied by birefringence.
The non-minimal CPT-even $\kI$ coefficients introduce
the possibility of nonbirefringent dispersion.
These give rise to a polarization-independent
speed that depends on even powers of energy.
Dispersion of this kind can be constrained
by looking for arrival-time differences
in explosive sources that produce photons
over a high range of energies, such as
GRBs.

\section{Resonant-cavity tests}

Astrophysical tests are primarily sensitive
to the vacuum coefficients.
Laboratory experiments complement these tests
and can access the vacuum-orthogonal coefficients.
Many laboratory experiments
rely on high-Q resonant cavities
and search for tiny direction- and
boost-dependent changes in
frequency that would indicate
Lorentz violation.\cite{cavities}

The change in frequency
can be determined perturbatively.
Given the conventional electromagnetic
fields $E$ and $B$ and vector potential $A_\mu$,
the shift in resonant frequency is approximated by
\beq
\fr{\de\nu}\nu =
-\fr{1}{4\vev{U}} \int d^3x \big(\vec E^*\cdot \de\vec D
-\vec B^*\cdot \de\vec H \big) \ ,
\label{dnu}
\eeq
where
\begin{align}
  \de\vec D &= \kde\cdot\vec E + \kdb \cdot\vec B + 2 \vec k_{AF} \times\vec A \ ,
  \notag \\
  \de\vec H &= \khe\cdot\vec E + \khb\cdot\vec B - 2(\kaf)_0 \vec A+ 2\vec k_{AF} A_0  \ .
\label{dfields}
\end{align}
The $\hat\ka$ matrices are
combinations of $(\kf)_{\ka\la\mu\nu}$ operators.
A problem arises when we apply this result
to the nonrenormalizable case.
The fields generally exhibit discontinuities
at one or more surfaces of the cavity.
The differential operators in Eq.\ \refeq{dfields}
acting on these fields then lead to divergences.
These divergences can be circumvented by
smoothly extending the fields in Eq.\ \refeq{dfields}
beyond the volume of the cavity.
Equation \refeq{dnu} then gives finite
shifts in resonant frequency in terms of
the coefficients for Lorentz violation.

While not as sensitive as astrophysical tests
are to vacuum coefficients,
cavity experiments can probe 
the much larger set of vacuum-orthogonal
coefficients.
Consequently,
these experiments are expected to
play an important role in constraining the
nonrenormalizable operators of the SME.

\end{document}